\documentstyle[epsfig,twocolumn]{jpsj}
\def\runtitle{Topological Excitations in Spinor Bose-Einstein Condensates}
\def\runauthor{Shunji {\sc Tuchiya} and Susumu {\sc Kurihara}}

\title
{
Topological Excitations in Spinor Bose-Einstein Condensates
}

\author
{ 
Shunji {\sc Tuchiya}\footnote{E-mail: tuchiya@kh.phys.waseda.ac.jp}
and Susumu {\sc Kurihara}}

\inst
{
Department of Physics, Waseda University, Tokyo 169-8555
}

\recdate
{
\today
}

\abst
{
We investigate the properties of skyrmion in the ferromagnetic state of spin-1 Bose-Einstein condensates by means of the mean-field theory and show that the size of skyrmion is fixed to the order of the healing length.
It is shown that the interaction between two skyrmions with oppositely rotating spin textures is attractive when their separation is large, following a unique power-law behavior with a power of -7/2.
}

\kword
{
Bose-Einstein condensation, spin degree of freedom, ferromagnetic state, skyrmion, spin texture
}

\begin{document}
\sloppy
\maketitle

Recent experimental success~\cite{rf:1} in obtaining Bose-Einstein condensation (BEC) of $ ^{23}{\rm Na}$ atoms in an optical trap has opened up new directions in the study of BEC.
Although confining atoms in a magnetic traps has been a successful method,
it aligns the orientation of spin due to the existance of strong magnetic field.
For this reason, spin of the trapped atoms cannot be regarded as a degree of freedom.
In contrast, optical trap is based on the optical dipole force.
Therefore atoms are confined in all hyperfine states.
This offers the possibility of studying condensates with a spin degree of freedom, the spinor condensates.

A variety of novel phenomena has been predicted for spinor condensates,
 such as changes in ground state structures with interaction parameters, propagation of spin waves and internal vortex structure.~\cite{rf:2,rf:3,rf:4} 
In particular, the possibility of creating vortex states without core, or skyrmions, in spinor condensates has been pointed out.~\cite{rf:2,rf:3}
Skyrmions, which do not have an ordinary vortex core due to the spin degree of freedom, offer us many new physical phenomena beyond those presented by other vortex states. 
One interesting feature is the reduction of kinetic energy associated with the rotation by transferring this energy to the spin and eliminating energetically unfavorable normal core.
This reduces the energy of skyrmion to a value lower than that of an ordinary singular vortex.

In this paper we investigate the properties of skyrmions in spin-1 Bose condensates.
First we discuss the basic properties of a single skyrmion, then we examine the interaction between two skyrmions with opposite phases.

{We shall consider bosons with hyperfine spin $F=1$. 
This includes alkali atoms with nuclear spin $I=3/2$ such as $ ^{23}{\rm Na}$, $ ^{39}{\rm K}$, and $ ^{87}{\rm Rb}$.
The order parameter of the Bose condensate is characterized by
\begin{eqnarray}
\langle\psi_\alpha({\mib r})\rangle&=&\Psi_\alpha({\mib r})\\ \nonumber
&=&\sqrt{n({\mib r})}\zeta_\alpha({\mib r}) \ \ (\alpha=1,0,-1) 
\end{eqnarray}
where $\alpha$ indicates three hyperfine components, $\psi_\alpha({\mib r})$ is the field annihilation operator of hyperfine state $|1,\alpha \rangle$, 
$n({\mib r})=\displaystyle \sum_\alpha|\Psi_\alpha({\mib r})|^2$, $\zeta_\alpha({\mib r})$ is a normalized spinor i.e. $\zeta^{\dagger}\cdot\zeta=1$.
This determines the average local spin
\begin{equation}
\langle{\mib F}\rangle\equiv\zeta_\alpha^\ast({\mib r}){\mib F}_{\alpha\beta}\zeta_\beta({\mib r}).
\end{equation}

Ho~\cite{rf:2} showed that in the case of spin-1 bosons consideration of two ground states is necessary since the effective interaction between two spins can be either ferromagnetic or antiferromagnetic.
The ground states of the former is minimized for $\langle{\mib F}\rangle^2=1$,
and the latter, which is known for the polar state, is minimized for $\langle{\mib F}\rangle=0$.
The properties of vortices are different between these two states, and skyrmions are possible only in the ferromagnetic state.
In this paper, we shall discuss their shape, size and interactions in the ferromagnetic state.

The energy of the condensate is
\begin{eqnarray}
K &=& \int {\rm d}^3r\biggl\{\frac{\hbar^2}{2M}({\nabla\sqrt{n}})^2 
+\frac{\hbar^2}{2M}(\nabla{\zeta})^{\dagger}(\nabla{\zeta})n \nonumber
\\&+&(U-\mu)n+\frac{1}{2}g_2n^2\biggr\},
\end{eqnarray}}
$\!\!$where $\mu$ is the chemical potential, $g_2=4\pi\hbar^2a_2/M$, $a_2$ is the $s$-wave scattering length for two colliding atoms with total spin 2, and $U$ is the trapping potential.
All spinors in the ferromagnetic state are related each other by gauge transformation $e^{i\theta}$ and spin rotations 
$R(\alpha,\beta,\tau)=e^{-iF_z\alpha}e^{-iF_y\beta}e^{-iF_z\tau}$,
where $(\alpha,\beta,\tau)$ are the Euler angles.
The spinor $\zeta$ is given by~\cite{rf:2}
\begin{eqnarray}
\zeta({\mib r})&=&e^{i\theta}R(\alpha,\beta,\tau)
\pmatrix{
1 \cr
0 \cr
0 \cr
}
\nonumber\\
 &=&
\pmatrix{
\frac{1}{2}e^{-i(\gamma+\alpha)}(1+\cos\beta) \cr
\frac{1}{\sqrt{2}}e^{-i\gamma}\sin\beta \cr
\frac{1}{2}e^{-i(\gamma-\alpha)}(1-\cos\beta) \cr
},
\end{eqnarray}
where $\gamma=\tau-\theta$.
The spin texture and superfluid velocity of the ferromagnetic state are given by
\begin{eqnarray}
\langle{\mib F}\rangle=(\sin\beta\cos\alpha)\hat{\mib x}+(\sin\beta\sin\alpha)\hat{\mib y}+(\cos\beta)\hat{\mib z},
\end{eqnarray}
\begin{eqnarray}
{\mib v}_{\rm s}=-\frac{\hbar}{M}(\nabla\gamma+\cos\beta\nabla\alpha).
\end{eqnarray}
Equation (5) shows that $\alpha$, $\beta$ characterizes the spin orientation. Moreover eq.(6) shows that spatial variation of spin in general leads to superflows.
Note that the spin texture with superfluid velocity takes the same form as the angular momentum texture with superfluid velocity in $^{3}{\rm He}$-A.~\cite{rf:6}


To illustrate the skyrmion, we use cyrindrical co-ordinates $(r, \phi, z)$ and assume that the spin texture and superfluid velocity of the condensate are both cylindrically symmetric.
Taking $\alpha=\phi$, $\gamma=-\phi$, $\beta=\beta(r)$, where $\beta(r)$ is a function increasing from $\beta=0$ at $r=0$. The spin texture and superfluid velocity field are
\begin{equation}
\langle{\mib F}\rangle=(\sin\beta\cos\phi)\hat{\mib x}
+(\sin\beta\sin\phi)\hat{\mib y}+(\cos\beta)\hat{\mib z}.
\end{equation}
\begin{equation}
{\mib v}_{\rm s}=\frac{\hbar}{Mr}(1-\cos\beta){\mib e}_\phi ,
\end{equation}
Equation (8) shows that the superfluid velocity vanishes continuously as $r\to0$, and energetically unfavourable normal core is avoided.~\cite{rf:2} For $\beta(r)=\pi/2$, the superfluid velocity of the skyrmion reduces to an ordinary singular vortex.

To obtain a precise order parameter of the skyrmion,
we have to determine $n(r)$ and $\beta(r)$ that minimize the energy.
To investigate only properties of the skyrmion we neglect the trapping potential.
From variational conditions $\delta K/\delta n=0$, $\delta K/\delta \beta=0$,
we can derive a pair of coupled equations for $n(r)$ and $\beta(r)$
\begin{eqnarray}
\frac{{\rm d}^2f}{{\rm d}x^2}+&\frac{1}{x}&\frac{{\rm d}f}{{\rm d}x}-\biggl\{\frac{1}{2}\biggl(\frac{{\rm d}\beta}{{\rm d}x}\biggr)^2
+\frac{1}{2x^2}(3-\cos\beta)(1-\cos\beta)\biggr\}f\nonumber \\
&-&(f^2-1)f=0 ,
\end{eqnarray}
\begin{equation}
\frac{{\rm d}^2\beta}{{\rm d}x^2}+\biggl(\frac{1}{x}+\frac{2}{f}\frac{{\rm d}f}{{\rm d}x}\biggr)\frac{{\rm d}\beta}{{\rm d}x}-\frac{1}{x^2}(2-\cos\beta)\sin\beta=0,
\end{equation}
where $x=r/\xi$ and $f=\sqrt{n/\bar{n}}$. Here we define $\xi=(\hbar^2/2Mg_2\bar{n})^{1/2}$ as the healing length of the condensate and $\bar{n}=\mu/g_2$ as the density of the condensate without flow.

In view of the absence of the core, we shall neglect the spatial dependence of density for the moment, and replace it with an average $\bar{n}$.
In this case, the solution of eq.(10) is
\begin{equation}
\beta(x)=2\tan^{-1}\left(\frac{4kx}{8-k^2x^2}\right).
\end{equation}
\vspace{5mm}
\begin{figure}
\vspace{1cm}
\begin{center}
\epsfig{file=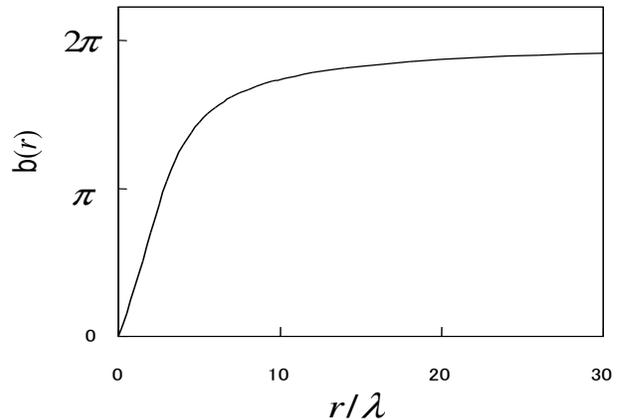,width=8cm,height=5.5cm}
\end{center}
\caption{The variational function $\beta(r)$ when the spatial dependence of density is neglected. $\beta(r)$ is a monotonically increasing function obeying $\beta(0)=0$ and $\beta(\infty)=2\pi$.}
\label{fig:1}
\end{figure}
We plot this function in Fig.~\ref{fig:1}.
Here we put $\beta^\prime(0)=k$.
$\vspace{-5mm}$
$\lambda=\xi/k$ expresses the typical size of the skyrmion.

The energy of skyrmion per unit length along the $z$-axis is
\begin{eqnarray}
E_{\rm s}
&=&\frac{4\pi\hbar^2\bar{n}}{M}
\frac{
\bigl(\frac{R}{\lambda}\bigr)^4
+\bigl(\bigl(\frac{R}{\lambda}\bigr)^4+64\bigr)
\tan^{-1}\bigl(\frac{1}{8}\bigl(\frac{R}{\lambda}\bigr)^2\bigr)}
{\bigl(\frac{R}{\lambda}\bigr)^4+64} \nonumber\\
&\sim&\left(1+\frac{\pi}{2}\right)\frac{4\pi\hbar^2\bar{n}}{M}\;\;\;\;\;(R\gg\lambda)
\end{eqnarray}
\begin{figure}
\begin{center}
\vspace{1cm}
\epsfig{file=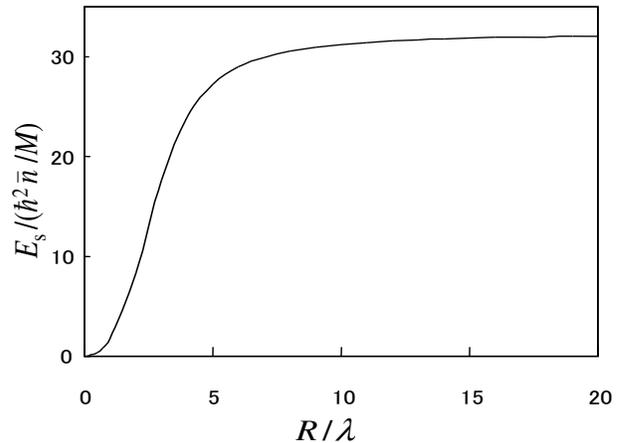,width=8cm,height=5.8cm}
\end{center}
\caption{The energy of the skyrmion in case of neglecting spatial dependence of the density. For $R\gg\lambda$, it approaches the constant $\left(1+\pi/2\right)4\pi\hbar^2\bar{n}/M$.}
\label{fig:2}
\end{figure}
where $R$ is the characteristic dimension of the condensate.
It is clear from eq.(12) that the skyrmion has less energy than that of a singular vortex which is logarithmically dependent on the size of system due to the existence of singularity at the core.
The $R$ dependence of the energy of the skyrmion is shown in Fig.~\ref{fig:2}.

Figure~\ref{fig:2} shows that the energy of the skyrmion is minimum when $\lambda$ is infinite. This result is in fact not unreasonable physically. Our previous assumption has neglected the spatial dependence of density, and only considerd spatial dependence of $\zeta$. Since $\zeta$ tends to vary as slow as possible in order to decrease the kinetic energy, the size of the skyrmion becomes infinitely large.
However as the size of the skyrmion increase, the region in which $n(\bf r)$ deviate from $\bar n$ becomes large. This increases the interaction energy.
Therefore considering spatial dependence of density, the balance between the kinetic energy and the interaction energy of the condensate determine the typical size of the skyrmion.

We can assume without loss of generality that deviation of $n(r)$ from the average density $\bar{n}$ is small i.e. $n(r)=\bar{n}(1-\epsilon(r))$, $\epsilon(r)\ll1$ for $r\to\infty$.
Linearizing eq.(9) with $\epsilon(r)$ and using eq.(11), we obtain the following equation
\begin{eqnarray}
\frac{{\rm d}^2\epsilon}{{\rm d}x^2}
&+&\frac{1}{x}\frac{{\rm d}\epsilon}{{\rm d}x}
-2\epsilon\nonumber\\
&=&-2\left(\frac{8k^3x^2+4^3k}{k^4x^4+4^3}\right)^2\left(1-\frac{1}{2}\epsilon\right).
\end{eqnarray}
The solution of the above equation has the following asymtotic behavior
\begin{eqnarray}
\epsilon(x)=
\left\{
\begin{array}{@{\,}ll}
C_1 e^{-\sqrt{2}\,x} & (x\gg1) \\
C_2I_0(\sqrt{k^2+2}\,x)+\dfrac{2k^2}{k^2+2} & (x\ll1),
\end{array}
\right.
\end{eqnarray}
where $I_0$ is a modified Bessel function.
Equation (14) shows that the variation of the density is spreaded over a distance $\xi$. Therefore the size of skyrmion is of the order of the healing length of condensate i.e. $\lambda\sim\xi$.
The skyrmion is thus stabilized by the fact that as it becomes large the gradient of spin texture on one hand decreases the kinetic energy and the variation of density increases the interaction energy on the other.

\begin{figure}
\begin{center}
\vspace{5mm}
\epsfig{file=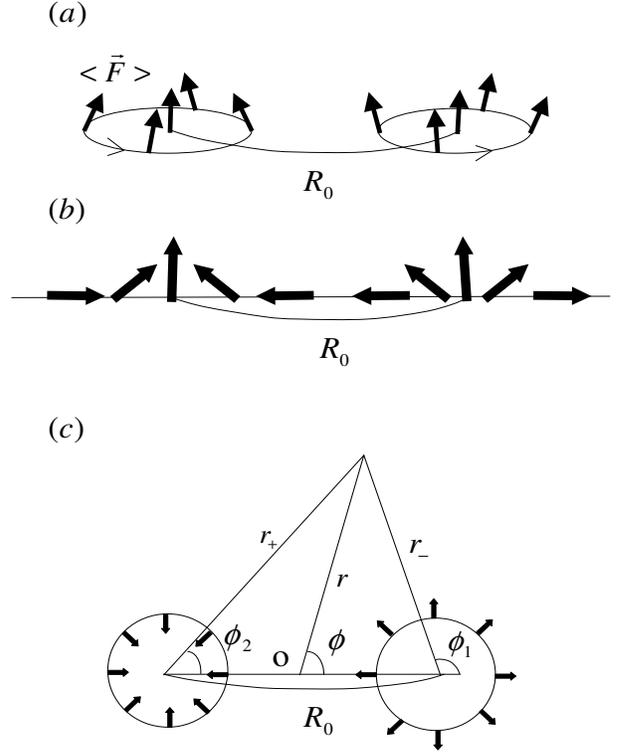,width=8cm,height=10cm}
\end{center}
\caption{(a) Schematic representation of two skyrmions. Bold arrows represent magnetic moment. The direction of circulation of each skyrmion is represented by a circular arrow. (b) Intersection of the spin texture. (c) View from the $z$ axis.}
\label{fig:3}
\end{figure}

Let us consider now two skyrmions with spin textures oppositely rotating
\begin{eqnarray}
\zeta_{\rm s}(r, \phi)
&=&\pmatrix{
\frac{1}{2}(1+\cos\beta(r)) \cr
\frac{1}{\sqrt{2}}e^{i\phi}\sin\beta(r) \cr
\frac{1}{2}e^{2i\phi}(1-\cos\beta(r))\cr
}\nonumber\\
\zeta_{\bar {\rm s}}(r, \phi)
&=&\pmatrix{
\frac{1}{2}(1+\cos\beta(r)) \cr
-\frac{1}{\sqrt{2}}e^{i\phi}\sin\beta(r) \cr
\frac{1}{2}e^{2i\phi}(1-\cos\beta(r))\cr
},
\end{eqnarray}
where $\beta(r)=2\tan^{-1}(\frac{4r/\lambda}{8-(r/\lambda)^2})$.
The minus sign in front of the second component of $\zeta_{\bar {\rm s}}$ represents opposite phase.
The texture in the presence of two skyrmions is schematically represented in Fig.~\ref{fig:3}.

We shall determine two-skyrmion energy as a function of their separation $R_0$.
Here we neglect the spatial dependence of density. The difference between the energy at a given separation $R_0$ and the energy of two isolated skyrmions
per unit length is
\begin{eqnarray}
E_{\rm two}(R_0) &=& \frac{\hbar^2}{2M}\bar{n}\int {\rm d}^2r(\nabla{\zeta_{\rm two}})^{\dagger}(\nabla{\zeta_{\rm two}})-2E_s.
\end{eqnarray}
For $R_0\gg\lambda$ we approximately express the order parameter in the presence of two skyrmions as a sum of each order parameter.
\begin{equation}
\zeta_{\rm two}(r, \phi)=\zeta_{\rm s}(r_-, \phi_1)+\zeta_{\rm \bar s}(r_+, \phi_2).
\end{equation}

\begin{figure}
\begin{center}
\vspace{5mm}
\epsfig{file=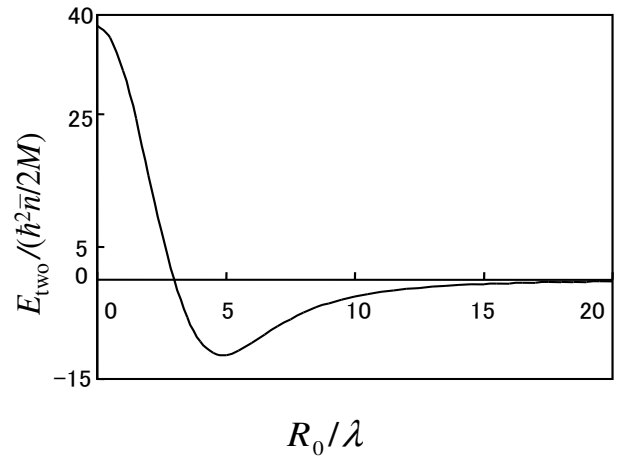,width=8cm,height=6cm}
\end{center}
\caption{The energy of two skyrmions as a function of the distance between the center of two skyrmions. }
\label{fig:4}
\end{figure}
\begin{figure}
\begin{center}
\vspace{5mm}
\epsfig{file=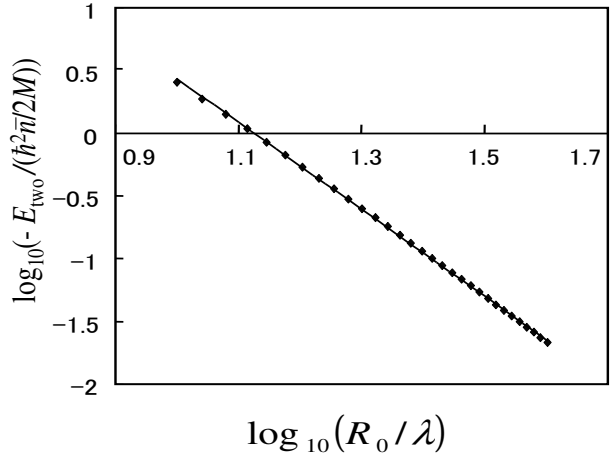,width=8cm,height=6cm}
\end{center}
\caption{Logarithmic plot of energy of two skyrmions. The solid line, with a slope of -7/2 is the best fit to the data.}
\label{fig:5}
\end{figure}

Our result is plotted in Fig.~\ref{fig:3}. The curve has a minimum at $R_0=r_0$ where $r_0$ is approximately $5\lambda$.
It can be seen that for $R_0<r_0$ the interaction between skyrmions is repulsive, while $R_0>r_0$ the interaction is attractive.
As two skyrmions are brought together the region where spin texture varies gradually overlap, thereby changes the kinetic energy.
Since texture tends to vary smoothly, skyrmions with large separation tends to come closer to each other, thus interaction becomes attractive. In contrast, if they are too closer together, the region where velocity increases become large and the energy of two skyrmion increases. Thus with sufficiently close separation the interaction energy is repulsive.

We have investigated numerically the attractive interaction between skyrmions. Figure~\ref{fig:5} shows logarithmic plot of energy of two skyrmions. It shows that attractive interaction between skyrmions is approximately proportional to $R_0^{-7/2}$.

In conclusion, we have investigated the energy and the size of a skyrmion in ferromagnetic state of spin-1 Bose-Einstein condensates and have shown that the size is of the order of the healing length. We also have shown that the skyrmion has lower excitation energy than that of a singular vortex. Therefore skyrmions can be excited more easily than singular vortices in scalar superfluids. This means that skyrmions are important in determining thermodynamic properties of the condensate at low temperatures. It also means a reduction of the critical superfluid velocity.

We have shown that the interaction between two skyrmions with opposite phases is attractive when their separation is larger than $r_0$ and repulsive when their separation is smaller than $r_0$. Physical origin and the power of attractive force between skyrmions is still unknown, and interactions between skyrmions with other textures is also an interesting subject to invetigate.

\section*{Acknowledgement}
The authors would like to thank M. Nishida for useful discussions.

\end{document}